\documentclass[journal]{IEEEtran}

\ifCLASSINFOpdf
\else
   \usepackage[dvips]{graphicx}
\fi
\usepackage{url}
\usepackage{cite}
\usepackage{amsmath}
\usepackage{amstext}
\usepackage{amsfonts}

\usepackage{graphicx}
\usepackage{color,soul}

\usepackage{framed}
\begin{document}
\onecolumn
\vspace*{8.5in}
\begin{framed}
This work has been submitted to the IEEE for possible publication. Copyright may be transferred without notice, after which this version may no longer be accessible.
\end{framed}
\twocolumn

\title{Personalizing Keyword Spotting with Speaker Information}

\author{Beltrán Labrador, \IEEEmembership{Student Member, IEEE}, Pai Zhu, Guanlong Zhao, \IEEEmembership{Member, IEEE}, Angelo Scorza Scarpati \newline Quan Wang, \IEEEmembership{Senior Member, IEEE}, Alicia Lozano-Diez, Alex Park, Ignacio López Moreno 
\thanks{B. Labrador and A. Lozano-Diez are with the Universidad Autónoma de Madrid, Madrid 28049, Spain (e-mail: beltran.labrador@uam.es). B. Labrador performed this work as an intern at Google and was partially supported by FPI RTI2018-098091-B-I00, MCIU/AEI/10.13039/501100011033/FEDER, UE and PID2021-125943OB-I00, MCIN/AEI/10.13039/501100011033/FEDER, UE from the Spanish Ministerio de Ciencia e Innovaci\'on, Agencia y del Fondo Europeo de Desarrollo Regional.}
\thanks{P. Zhu, G. Zhao, A. S. Scarpati, Q. Wang, A. Park, and I. L. Moreno are with Google LLC, New York, NY 10011 USA (e-mail: paizhu@google.com).}}

\maketitle

\begin{abstract}

Keyword spotting systems often struggle to generalize to a diverse population with various accents and age groups. To address this challenge, we propose a novel approach that integrates speaker information into keyword spotting using Feature-wise Linear Modulation (FiLM), a recent method for learning from multiple sources of information. We explore both Text-Dependent and Text-Independent speaker recognition systems to extract speaker information, and we experiment on extracting this information from both the input audio and pre-enrolled user audio. We evaluate our systems on a diverse dataset and achieve a substantial improvement in keyword detection accuracy, particularly among underrepresented speaker groups. Moreover, our proposed approach only requires a small 1\% increase in the number of parameters, with a minimum impact on latency and computational cost, which makes it a practical solution for real-world applications.

\end{abstract}

\begin{IEEEkeywords}
Keyword spotting, target speaker personalization, AI fairness
\end{IEEEkeywords}

\IEEEpeerreviewmaketitle

\section{Introduction}

\IEEEPARstart{K}{eyword} spotting (KWS) is the task of detecting specific words or phrases from an audio stream. This technology is widely used in different applications such as speech data mining, audio indexing, and wake-up word detection, which is often used to initiate an interaction with voice assistants in mobile phones, smart speakers, and other internet of things (IoT) devices (e.g. ``Okay Google", ``Hey Siri", or ``Alexa") \cite{kws_overview, kws_apple, kws_end2end, alexa_shi19c_interspeech}.

A robust KWS system is expected to generalize to queries that may contain diverse accents, different age groups, and varying acoustic environments. A straightforward solution is to increase the KWS model capacity and train it with a diverse dataset. However, in many use cases, keyword spotting applications are intended for low-resource devices that are characterized for having memory, computation and power constraints. This requires KWS systems to be lightweight with a relatively small memory and power footprint \cite{Sun2017,raw_dnn,dnn_handling,cnn_handling,kws_rnn_1, kws_snn, kws_maxpooling, park21_interspeech, tt_kws}. Therefore, parameter-efficient modeling strategies are favored over simply increasing the model size.

Many voice products provide interfaces to allow users to enroll their voice identity~\cite{wang2020version} to customize the user experience. The enrollment process generally takes place in the form of computing speaker embedding vectors from user-provided enrollment utterances. A speaker recognition model can be used to compute these speaker embeddings such as i-vectors, x-vectors, or d-vectors \cite{ivectors, xvectors, td_sid}.

Using auxiliary speaker information is a common approach to adapt neural systems to different speaker characteristics. For instance, in the Automatic Speech Recognition (ASR) task, feature space maximum likelihood linear regression (fMLLR) features have been shown to outperform traditional features (e.g. MFCC) thanks to the speaker adaptation process \cite{Parthasarathi2015}. ASR systems have also been directly conditioned with speaker embeddings \cite{baskar22b_interspeech}. These speaker embeddings can also be used to improve the performance of Voice Activity Detection (VAD) systems \cite{ding20_odyssey,ding22_interspeech}. These so-called ``Personal VAD" systems typically use a small neural network to predict whether a given frame of audio contains speech, and if the speech is produced by a enrolled speaker or a non-enrolled speaker. The speaker embedding information is used to condition the neural network on the target speaker's identity, which can help the network to learn more discriminative features for the target speaker's voice on a computational constricted scenario.

For the keyword spotting task there has been some previous efforts on learning to customize the keyword detection to a target speaker. In \cite{kws_with_targetspeaker}, the authors use a speech enhancement frontend based on VoiceFilter \cite{wang19h_interspeech} to perform speech extraction in order to only feed the target speaker audio to the keyword spotting system. Furthermore, in \cite{yang22l_interspeech}, by leveraging speaker information, multi-task learning has been applied to jointly train a system to discriminate between keywords and speakers, and afterwards perform a task specific adaptation. However, this multi-task approach may suffer from the inefficiency as a result of trying to address two tasks simultaneously, as well as reduced model interpretability and increased computational complexity \cite{caruana1997multitask}.  

This work is inspired by the findings in \cite{locale_encoding}, where the authors show that by conditioning a small KWS system on different locale indices they were able to improve the system's performance in multilingual contexts. Thus, we propose to use speaker embeddings extracted from a pretrained speaker recognition system to condition a KWS system using Feature-wise Linear Modulation (FiLM) \cite{film_paper}, an effective method to integrate and understand multiple sources of information. This allows the model to adapt to different speaking styles while keeping the restrictive computational and memory constraints.

In our experiments, we compare the performance when conditioning the KWS detection using Text-Dependent (TD) speaker embeddings with respect to employing Text-Independent (TI) representations. We analyze the results on a diverse dataset, considering various locales and age groups. Additionally, we introduce a robust training strategy to improve the system's suitability for production scenarios.

The rest of this paper is organized as follows. Section \ref{sec:method} introduces the baseline KWS system, the speaker recognition models, as well as how we personalize keyword detection. Section \ref{sec:experimental} details the experimental setup including data preparation, system descriptions, and training configurations. Section \ref{sec:results} presents the results for different models and population groups. Finally, we conclude the paper in Section \ref{sec:conclusion}.

\section{Methods}
\label{sec:method}

\subsection{Baseline}

Our baseline is an end-to-end neural KWS system optimized for low-resource use cases, as described in \cite{kws_end2end}. 
This model has an encoder-decoder architecture with a total of 350K parameters, where the encoder has 4 Singular Value Decomposition Filter (SVDF) layers \cite{svdf} with 576 nodes in each layer and a memory of 6 frames. Each SVDF layer is followed by a bottleneck layer of size 64. The decoder has three SVDF layers with 32 nodes each and a memory of 32 frames.
This system \textit{f} is trained with examples composed of feature sequences (denoted as \textit{X}) and corresponding label sequences (denoted as \textit{Y}) that identify the keyword frames. We minimize the cross-entropy (CE) loss by finding optimal trainable parameters $\theta_{base}$ for the keyword detection task. The training dataset is mixed with all speakers \textit{S}.

\begin{equation}
\label{baseline_equation}
\begin{aligned}
\theta_{base}=& \ \mathop{\arg\!\min}_{\theta} \ \mathbb{E}(x, y) [\textit{L}_{CE}(f(x; \theta), y)] \\
& \ \text{where}\quad (x, y) \in \bigcup_{s=1..S} (X_s , Y_s)
\end{aligned}
\end{equation}
we use $\mathbb{E(*)[...]}$ to denote expectation over \textit{$*$}. More details of this baseline can be found in \cite{kws_end2end}.

\begin{figure}[t]
\centering
\includegraphics[trim={0in 0in 0in 0in},clip,width=3.4in]{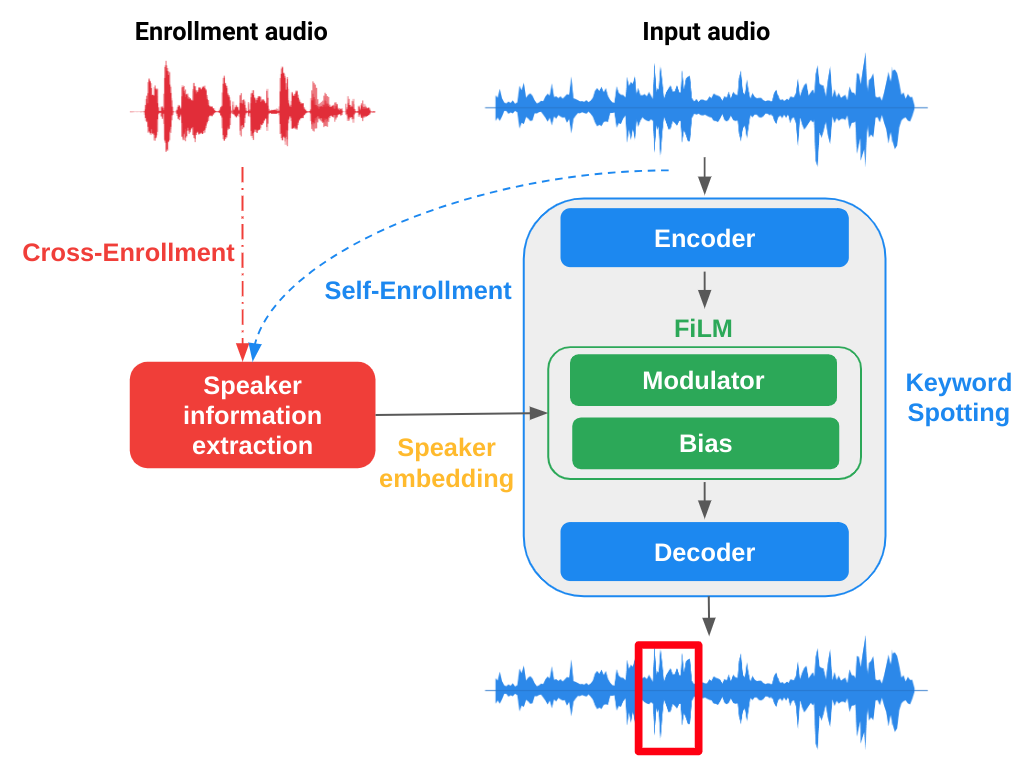}
\caption{We personalize the keyword detection with target speaker information applying FiLM over the encoder output logits with a speaker embedding. In the Self-Enrollment approach, this embedding vector is extracted directly from the input audio, while in the Cross-Enrollment strategy it is extracted from an enrollment audio spoken by the same individual.}
\label{fig_diagram}
\end{figure}

\subsection{Speaker Embedding}

We employ two distinct speaker verification systems to extract speaker embeddings. The first of these systems is a Text-Dependent (TD) system  \cite{td_sid}, which imposes the requirement that the speaker utters a specific phrase, in our case, ``Okay/Hey Google".
We also experiment with a Text-Independent (TI) system , which operates without any speech constraint, allowing to perform speaker verification on spontaneous, free speech. 

The TD system \cite{td_sid} has a compact memory footprint of 235k parameters. To train this system, we extract speech segments containing only the target keyword, a task facilitated by a pretrained keyword spotting system. This TD system consists of three Long Short-Term Memory (LSTM) layers with projection, each with 128 memory cells. A linear transformation layer follows the final projected LSTM layer to produce a 64-dimensional speaker embedding.

For the Text-Independent speaker encoder, we employ a conformer system based on the one described in \cite{wang2022highly} (Section 2.3.1), consisting of 12 conformer \cite{gulati2020conformer} encoder layers each of 256 dimensions, followed by an attentive temporal pooling mechanism \cite{wang22b_odyssey}, with a total of 22 million parameters, trained with the generalized end-to-end extended-set softmax (GE2E-XS) loss \cite{td_sid,pelecanos21_interspeech}. We use this system to extract a 256-dimensional speaker embedding.

\subsection{Personalizing detection with speaker embedding}

To condition the keyword detection with speaker information, we incorporate Feature-wise Linear Modulation (FiLM) \cite{film_paper} into our framework as shown in Fig. \ref{fig_diagram}. Using FiLM we can apply affine transformations to intermediate layer outputs within the neural network. These transformations serve as a mean of dynamically adjust the intermediate layer outputs, allowing the model to adapt its representations based on the unique characteristics of each speaker, as captured by the speaker embedding. 

\begin{equation}
\label{film_equation}
\textrm{FiLM}(l, \gamma, \beta) = \gamma \odot l + \beta
\end{equation}
where $\odot$ denotes element-wise multiplication.

The FiLM mechanism effectively learns  
scaling and bias functions ($\gamma$ and $\beta$ in equation \ref{film_equation}), which are integrated as trainable projection layers connecting the speaker embedding. We apply this modulation to the encoding logits (\textit{l} in equation \ref{film_equation}): the layer output between the encoder and decoder networks. This allows the model to perform the keyword detection task with awareness of speaker characteristics.
\begin{equation}
\label{train_cond_equation}
\begin{aligned}
\theta_{cond}=& \ \mathop{\arg\!\min}_{\theta} \ \mathbb{E}(x, y, s) [\textit{L}_{CE}(f(x, s; \theta), y)] \\
& \ \text{where}\quad (x, y) \in (X_s, Y_s), s \in S
\end{aligned}
\end{equation}

We update the parameters $\theta_{cond}$ by minimizing the cross-entropy loss, given feature sequences $X_s$, label sequences $Y_s$ and conditioned by the enrollment speaker embeddings stored in \textit{S}.

\begin{table*}
    \centering
    \caption{Performance in EER (\%) of the systems on different population groups (English locales/accents \& age)}
\begin{tabular}{c|cc|cc|cc|cc|cc}
                          & \multicolumn{2}{c|}{All locales ~} & \multicolumn{2}{c|}{India ~} & \multicolumn{2}{c|}{United States ~} & \multicolumn{2}{c|}{Great Britain ~} & \multicolumn{2}{c}{Australia}                                  \\
\textbf{EER (\%)}                  & All ages & \textless12                      & All ages & \textless12                  & All age & \textless12                  & All ages & \textless12                  & All ages & \textless12                                                \\ 
\hline
Baseline                  & 1.93\%  & 3.59\%                  & 3.31\%  & 3.7\%               & 3.74\%  & 4.98\%              & 0.6\%   & 1.85\%              & 0.78\%  & \begin{tabular}[c]{@{}c@{}}1.07\%\\\end{tabular}  \\
Text-Indep. Self-Enrollment        & 1.57\%  & 3.01\%                  & 2.52\%  & 3.2\%               & 3.41\%  & 3.96\%              & 0.51\%  & 1.66\%              & 0.73\%  & 0.78\%                                            \\
Text-Indep. Cross-Enrollment       & 2.12\%  & 3.75\%                  & 3.81\%  & 2.88\%              & 3.85\%  & 5.18\%              & 0.6\%   & 1.7\%               & 0.71\%  & 0.99\%                                            \\
Text-Dep.                        & 1.88\%  & 3.38\%                  & 3.26\%  & 2.79\%              & 3.75\%  & 4.43\%              & 0.63\%  & 1.59\%              & 0.67\%  & 0.84\%                                            \\
\textbf{Relative improvement (\%)} & ~       & ~                       & ~       & ~                   & ~       & ~                   & ~       & ~                   & ~       & ~                                                 \\ 
\hline
Text-Indep. Self-Enrollment vs Baseline             & \textbf{-18.7\%} & \textbf{-16.2\%}                 & \textbf{-23.9\%} & -13.5\%             & \textbf{-8.8\%}  & \textbf{-20.5\%}             & \textbf{-15.0\%} & -10.3\%             & -6.4\%  & \textbf{-27.1\%}                                           \\
Text-Indep. Cross-Enrollment vs Baseline            & 9.8\%   & 4.5\%                   & 15.1\%  & -22.2\%             & 2.9\%   & 4.0\%               & 0.0\%   & -8.1\%              & -9.0\%  & -7.5\%                                            \\
Text-Dep. vs Baseline            & -2.6\%  & -5.9\%                  & -1.5\%  & \textbf{-24.6\%}             & 0.3\%   & -11.0\%             & 5.0\%   & \textbf{-14.1\%}             & \textbf{-14.1\%} & -21.5\%                                          
\end{tabular}
    \label{table:full_results}
\end{table*}

\section{Experimental setup}
\label{sec:experimental}

In all experiments, we use the same train and evaluation datasets, front-end features, and data augmentations. 
When handling user data, we abide by Google's AI principles \cite{aiprinciples} and privacy principles \cite{privacyprinciples}.

\subsection{Data description and preparation}

The training and evaluation datasets consist of vendor-provided data. No user data were used in these experiments. Text prompts were provided to vendors, who recorded their spoken utterances based on the given transcripts. The subset that contains the targeted keyword (``Okay/Hey Google") are referred as the positive dataset and otherwise the negative dataset. We divided the collected datasets into training, development and evaluation datasets, with no overlapping speakers. Moreover, the training set has been augmented with different transformations using room impulse simulations and varying degrees of noise and reverberation, producing 25 augmented copies for each original utterance \cite{kim2017generation}.

Our dataset comprises diverse English accents (including those from the US, India, UK, and Australia) and a spectrum of acoustic conditions (e.g., recordings from inside vehicles and various background noise scenarios). It exhibits a balanced gender distribution and it is meticulously designed to maintain equal proportions of near-field and far-field audio recordings. 

For these experiments, we extend the datasets by introducing enrollment data, where each utterance is augmented by pairing it with a corresponding same-speaker positive enrollment utterance.
As we construct these datasets, our aim is to simulate the conditions of a production environment where users typically enroll their voices during their initial set up with the device, resulting in the extraction of speaker embeddings from enrollment utterances.

\subsection{System description}

We evaluate the performance of three different approaches, depending on the speaker embedding types and how we simulate the enrollment utterances:

\begin{itemize}
\item
\textbf{Text-Independent Self-Enrollment}: In this setup, we extract a Text-Independent speaker embedding from the same utterance used as input for the keyword detection. 
This implies that the speaker information that conditions the KWS encoder's output, is extracted from the exact same speech sample in which we aim to identify the target keyword.
\item \textbf{Text-Independent Cross-Enrollment}: Here, the speaker embedding is extracted from a simulated enrollment utterance (by choosing a random utterance from the same speaker). This setup is closer to a production environment, where the enrollment utterances in production are pre-enrolled and are different from the query utterance.

\item  \textbf{Text-Dependent Cross-Enrollment}: We use a Text-Dependent (TD) speaker embedding from the simulated enrollment utterance mentioned above. To obtain the TD speaker embedding, we first use a pretrained keyword spotting model to extract the segment of the enrollment utterance where the keyword is spoken. Therefore, in this approach, the utterance segment that is used to compute the speaker embedding is constrained to the keyword, thus containing less variability and noise.

\end{itemize}

Note that the Text-Dependent Self-Enrollment scenario is not viable in our setup. This limitation comes from the absence of keyword in negative utterances, and the restriction of Text-Dependent SID systems to extract the speaker embedding from the target keyword segment.

\subsection{Training a robust model for non-enrollment conditions}
\label{subsec:robust_training}
In a real-world scenario, enrollment utterances may not always be available due to a failed or skipped enrollment stage. It is essential that the model can still accurately detect the keyword without relying on the speaker's enrollments. To ensure model's robustness, we have trained a model mixing utterances containing the speaker embedding with utterances without any speaker embedding.
Specifically, we randomly replace the enrollment utterance speaker embedding with a same dimension constant vector. This forces the model to learn to detect the keyword both with and without the speaker embedding information.

\section{Results and discussion}
\label{sec:results}

In this section, we present the results of the keyword detection task, focusing on the Equal Error Rate (EER) as single value metric and Detection Error Tradeoff (DET) curves showcasing model performance at various operating points.

\begin{figure}[t!]
\centering
\includegraphics[trim={0in 0in 0.35in 0.35in},clip,width=3.4in]{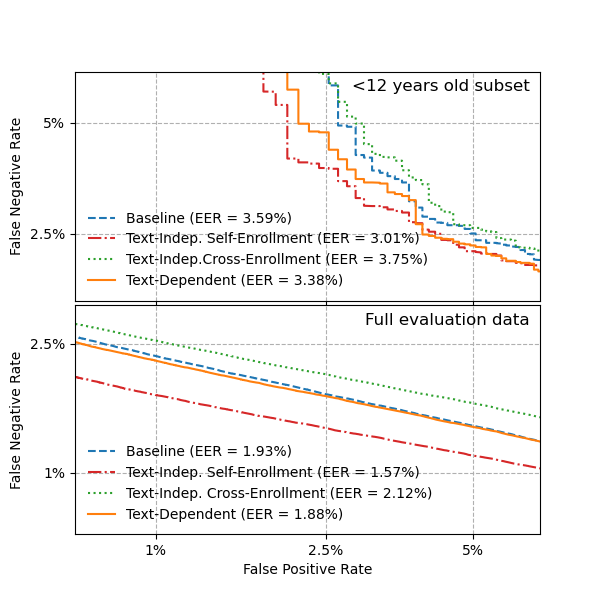}
\caption{Detection Error Trade-off (DET) curves illustrating the performance of the different approaches on the full evaluation data and on the under 12 years old evaluation subset.}
\label{fig_systems_alldata}
\end{figure}

Table \ref{table:full_results} and Fig. \ref{fig_systems_alldata} summarize the main results of these experiments. Specifically, we evaluate the three variations mentioned before: Text-Independent Self-Enrollment, Text-Independent Cross-Enrollment, and Text-Dependent systems. For each of these systems, we report the EER and DET curves across the entire dataset, as well as stratified by different locales and age groups (all ages and those under 12 years old). Furthermore, we show the relative improvement over the baseline for each configuration.

We observe that the Text-Independent Self-Enrollment approach yielded the most substantial improvement over the baseline model, achieving an 18\% relative improvement in terms of EER. This underscores the potential of the proposed strategy, as it leverages speaker embedding extracted from the same target utterance. However, it is crucial to note that this approach introduces elevated computational demands and system latency at inference time, due to the substantial increase of operations within the combined network. Therefore, it may be less practical for production environments, specially those with strict real-time processing requirements. 

Conversely, in the Text-Independent Cross-Enrollment scenario, where speaker embeddings are derived from enrollment utterances, we generally observed a degradation in performance. This result can likely be attributed to the high dimensionality of the 256-dimensional embedding, which may underfit the single layer FiLM. Furthermore, this embedding built on top of diverse utterances may seem less effective at capturing speaker characteristics as compared to building an embedding from just target keyword segments.

Therefore, in the Text-Dependent scenario, where a 64-dimensional speaker embedding is extracted from the enrollment utterance keyword segment, is expected to have a more reliable representation of the speaker characteristics. This approach has a notable 2.6\% relative improvement over the baseline, a 5.9\% EER relative improvement in children data, and as far as a remarkable 24\% enhancement when dealing with children data in the distinctly accented India locale. Importantly, the system barely adds any additional computation or latency in the production environment, as the speaker embeddings are pre-computed during the enrollment stage when the devices are first set up.

\begin{table}[t!]
    \centering
    \caption{Performance of the robust training for the Text-Dep. approach explained in subsection \ref{subsec:robust_training} on different speaker enrollment conditions}
    \begin{tabular}{ccc}
    \hline
     & With spk. embedding & Without spk. embedding\\  \hline
    Baseline &  - & 1.93 \% \\
    Text-Dep. & 1.88 \% & 39.54 \% \\
    Robust Text-Dep. & 1.85 \% & 2.03 \% \\
    \hline
    \end{tabular}
    
    \label{table:robust}
    \vspace{4pt}
\end{table}

Finally, we compare enrollment and non enrollment scenarios (with no pre-enrolled utterance) to assess the impact of enrollment utterances on the keyword spotting performance and underscore the importance of a robust model. As shown in Table \ref{table:robust}, in the absence of enrollment data, the TD-conditioned keyword spotting system experiences a complete failure. However, with our robust training approach (Section \ref{subsec:robust_training}), we ensure that the system remains consistent with the baseline performance. Furthermore, when provided with an enrollment speaker embedding, it yields a remarkable 4.1\% EER relative improvement over the baseline, enhancing also the regular TD approach by introducing a regularization effect, showing the adaptability of our approach, particularly in situations where speaker enrollment data is not present.

\section{Conclusion}
\label{sec:conclusion}
In this paper, we propose a novel method to personalize keyword spotting by integrating speaker information through FiLM modulation. We compare various approaches and speaker embedding types to identify the most effective strategies, demonstrating notable performance improvement, specifically when handling non-standard speech patterns and diverse speaker profiles.
This approach holds the potential to benefit a wide range of applications, by ensuring that the technology can effectively understand and respond to the unique voices and needs of diverse individuals, making the technology more accessible, adaptive, and inclusive.

\section*{Acknowledgment}
The authors thank  Joaqu\'in Gonz\'alez Rodr\'iguez, Doroteo Torre Toledano, Jacob Bartel, Andre Perunicic, Daniel Ramos, Fran\c{c}oise Beaufays, and Pedro Moreno Mengibar for their help.

\bibliographystyle{IEEEbib}
\bibliography{refs}

\end{document}